# Multiexciton generation in IV-VI nanocrystals: The role of carrier effective mass, band mixing, and phonon emission

Gal Zohar,[(a)] Roi Baer,[(b)] and Eran Rabani[(c)]

*(a) School of Physics and Astronomy, The Raymond and Beverly Sackler Faculty of Sciences, Tel Aviv University, Tel Aviv 69978 Israel. (b) Fritz Haber Center for Molecular Dynamics, Chaim Weizmann Institute of Chemistry, The Hebrew University of Jerusalem, Jerusalem 91904 Israel; (c) School of Chemistry, The Raymond and Beverly Sackler Faculty of Sciences, Tel Aviv University, Tel Aviv 69978 Israel.*

Abstract: We study the role of the effective mass, band mixing and phonon emission on multiexciton generation in IV-VI nanocrystals. A 4-band $k \cdot p$ effective mass model, which allows for an independent variation of these parameters, is adopted to describe the electronic structure of the nanocrystals. Multiexciton generation efficiencies are calculated using a Green's function formalism, providing results that are numerically similar to impact excitation. We find that multiexciton generation efficiencies are maximized when the effective mass of the electron and hole are small and similar. Contact with recent experimental results for multiexciton generation in PbS and PbSe is made.

## I. INTRODUCTION

The study of multiexciton generation (MEG) in nanocrystals (NC) has received considerable attention in recent years, driving experiments[1-18] and theory.[14, 19-35] While early stages have led a controversy over efficiencies of MEG in confined systems, over the last several years there is a consensus of efficiencies of 20-30% at $3E_g$ (where $E_g$ is the confined band gap),[23] depending on the size and composition of the NCs.

Despite advances made in the understanding of the MEG process, there are still many open questions. One of the most significant issues is related to the role of the effective mass of the carriers on the MEG efficiencies. On the one hand, the common wisdom suggests that a large ratio of effective masses should favor MEG, since this would lead to asymmetric excitation where the lighter particle takes most of the excess photon energy, thereby, reducing the threshold of MEG. On the other hand, materials such as PbS exhibit large MEG efficiencies where the effective masses of the electron and hole are very similar.[3, 18, 36]

In this paper we address the role of the effective mass of the electron and hole on the MEG process. We resort to a simple 4-band $k \cdot p$ effective mass model[37] which provides means to modify independently the effective mass of each charge carrier, preserving the remaining physical parameters. In addition, we explore the role of band mixing and the effect of phonon emission rate on MEG efficiencies.

The structure of this papers is as follows: In Section II we briefly present the theory of MEG based on the approach detailed in Ref. 26. The electronic structure approach described within a $k \cdot p$ effective mass model is presented in Section III. Results and conclusions are given in Sections IV and V, respectively.

## II. MEG THEORY

We follow the approach detailed in Ref. 26 to describe the MEG process. The electronic Hamiltonian can be partitioned as follows:

$$H = H_0 + H_{ph} - \mathcal{E}\mu \sin(\omega t), \quad (1)$$

where $H_0$ is the unperturbed Hamiltonian of the various excitonic states and their Coulomb interactions:

$$H_0 = \begin{pmatrix} E_0 & 0 & W_{0B} & \cdots \\ 0 & H_S & W_{SB} & \cdots \\ W_{0B}^\dagger & W_{SB}^\dagger & H_B & \cdots \\ \vdots & \vdots & \vdots & \ddots \end{pmatrix} \quad (2)$$

with $E_0$ the ground state energy of $H_0$. $H_S$ and $H_B$ are the Hamiltonian matrices of the single exciton subspace and biexciton subspace, respectively. Higher multi-exciton states are ignored. We assume, as in Hartree-Fock theory, that the coupling of the ground state to any singly excited state vanishes, $W_{0S} = 0$ (Brillouin's theorem). As commonly assumed in solid state theory, we neglect the contribution of higher excitons to the ground state, i.e. $W_{0B} \approx 0$. $W_{SB}$ describes the couplings between single excitons and biexcitons[26]:

$$W_{SB} = \delta_{ac} \left[ V_{jikb} - V_{kijb} \right] + \delta_{ab} \left[ V_{kijc} - V_{jikc} \right] + \delta_{ij} \left[ V_{kcab} - V_{ackb} \right] + \delta_{ki} \left[ V_{acjb} - V_{jcab} \right], \quad (3)$$

with the Coulomb matrix elements defined by:

$$V_{rsut} = \frac{1}{\epsilon} \iint d^3r\, d^3r' \left[ \frac{\psi_r^*(\mathbf{r})\psi_s^*(\mathbf{r})\psi_u(\mathbf{r'})\psi_t(\mathbf{r'})}{|\mathbf{r}-\mathbf{r'}|} \right]. \quad (4)$$

In the above equation, $\psi_r(\mathbf{r})$ are the single particle spin-orbitals and $\epsilon$ is the dielectric constant of the NC (assumed independent of $\mathbf{r}$ and $\mathbf{r'}$). We also do not assume spin degeneracy of the orbitals, since, as discussed below, the wave functions for type IV-VI materials are not spin-degenerate.

In Eq. (1), $H_{ph}$ represents the Hamiltonian of the phonons but in the sequel the electron-phonon interaction will be incorporated in a phenomenological way. The coupling to the electromagnetic field is described by the term $\mathcal{E}\mu \sin(\omega t)$ where:

$$\mu = \begin{pmatrix} 0 & \mu_{0S} & 0 & \cdots \\ \mu_{0S}^{\dagger} & \mu_S & \mu_{SB} & \cdots \\ 0 & \mu_{SB}^{\dagger} & \mu_B & \cdots \\ \vdots & \vdots & \vdots & \ddots \end{pmatrix}. \quad (5)$$

To obtain the MEG efficiency, we adopt the Green's function formalism discussed in Ref. 26 within the semi wide - band limit. The rate for transition into single- and bi-excitonic states following absorption of a photon of frequency $\omega$ is given by:

$$r_S(\omega) = \frac{\mathcal{E}^2}{\hbar} \sum_S \frac{\gamma |\mu_{0S}|^2}{(E_0 + \hbar\omega - E_S)^2 + (\gamma + \Gamma_S)^2 / 4}$$

$$r_B(\omega) = \frac{\mathcal{E}^2}{\hbar} \sum_S \frac{\Gamma_S |\mu_{0S}|^2}{(E_0 + \hbar\omega - E_S)^2 + (\gamma + \Gamma_S)^2 / 4}, \quad (6)$$

and the number of generated excitons by:

$$n_{ex}(\omega) = \frac{\sum_S \frac{(\hbar\gamma + 2\hbar\Gamma_S)|\mu_S|^2}{(E_0 + \hbar\omega - E_S)^2 + (\hbar\gamma + \hbar\Gamma_S)^2 / 4}}{\sum_S \frac{(\hbar\gamma + \hbar\Gamma_S)|\mu_S|^2}{(E_0 + \hbar\omega - E_S)^2 + (\hbar\gamma + \hbar\Gamma_S)^2 / 4}}, \quad (7)$$

where $\gamma$ is the phonon decay rate, assumed independent to energy, $\Gamma_S = \frac{2\pi}{\hbar} \sum_B |W_{SB}|^2 \delta(E - E_B)$ is the rate of the decay of a single exciton $|S\rangle$ to biexcitons,[26] $E_0$, $E_S$ and $E_B$ are the ground-state, singly and doubly excited state energies, respectively. $\mu_S$ is the transition dipole between the ground state and the singly excited state $|S\rangle$ where the hole (electron) is in state $|\psi_i\rangle$ ($|\psi_a\rangle$):

$$\mu_S^2 = \frac{1}{3} \sum_i |\langle \psi_a | \mathbf{e}_i \cdot \mathbf{p} | \psi_i \rangle|^2. \quad (8)$$

Here, $\mathbf{e}_i$ is the unit vector representing the direction of light polarization and $\mathbf{p} = -i\hbar\nabla$ is the momentum operator.

## III. 4-BAND EFFECTIVE MASS MODEL

We adopt a 4-band effective mass model developed by Kang and Wise[37] for calculating eigenstates and eigenenergies for the IV-VI nanocrystals. The Hamiltonian in the spherical approximations is given by:

$$H_0(\mathbf{k}) = \begin{pmatrix} \left(\frac{E_g^{bulk}}{2} + \frac{\hbar^2 k^2}{2m_e}\right)\mathbf{1} & \frac{\hbar P}{m_0} \mathbf{k} \cdot \boldsymbol{\sigma} \\ \frac{\hbar P}{m_0} \mathbf{k} \cdot \boldsymbol{\sigma} & -\left(\frac{E_g^{bulk}}{2} + \frac{\hbar^2 k^2}{2m_h}\right)\mathbf{1} \end{pmatrix} \quad (9)$$

where $E_g^{bulk}$ is the bulk band gap, $P$ coupling parameter between the valance and conduction bands, $m_0$ is the electron mass, $m_h$ and $m_e$ are the effective masses of the hole and electron, respectively, and is $\boldsymbol{\sigma}$ are Pauli matrices. In

Table 1: 4-band effective mass model parameters from PbS and PbSe. we provide the values of these parameters for PbS and PbSe.

Table 1: 4-band effective mass model parameters from PbS and PbSe.

|  | $E_g^{bulk}$ | $m_0/m_e$ | $m_0/m_h$ | $2P^2/m_0$ | $\epsilon$ |
|---|---|---|---|---|---|
| PbS | 0.41 eV | 2.5 | 3.0 | 2.5 eV | 17 |
| PbSe | 0.28 eV | 3.9 | 6.9 | 2.6 eV | 23 |

The spin-orbital is represented as a sum over each element of the vector wave function multiplied by the appropriate band-edge Bloch function $u_s(\mathbf{r})$:

$$\psi_i(\mathbf{r}) = \sum_{s=1}^{4} \phi_s^i(\mathbf{r}) u_s(\mathbf{r}) \quad (10)$$

where $i \equiv (n, m, j, \pi)$ is a composite index depending on the parity and angular momentum of the state. The boundary conditions $\phi_s^i(R) = 0$, where $R$ is the radii of the nanocrystals, are imposed, corresponding to approximating the confinement potential as an infinite step function. The above model Hamiltonian has an exact solution given by:[37]

$$\phi^{(n,j,m,(-1)^{l+1})} = \frac{N_{n,j,m,(-1)^{l+1}}}{R^{3/2}} \begin{pmatrix} i\left[j_l(k_\pm r) + \frac{j_l(k_\pm R)}{i_l(\lambda_\pm(k)R)} i_l(\lambda_\pm(k)r)\right] \begin{pmatrix} \sqrt{\frac{l+m+1/2}{2l+1}} Y_{l,m-1/2}(\theta,\varphi) \\ -\sqrt{\frac{l-m+1/2}{2l+1}} Y_{l,m+1/2}(\theta,\varphi) \end{pmatrix} \\ -\rho_\pm(k_\pm)\left[j_{l+1}(k_\pm r) + \frac{j_{l+1}(k_\pm R)}{i_{l+1}(\lambda_\pm(k)R)} i_{l+1}(\lambda_\pm(k)r)\right] \begin{pmatrix} \sqrt{\frac{l-m+3/2}{2l+1}} Y_{l+1,m-1/2}(\theta,\varphi) \\ \sqrt{\frac{l+m+3/2}{2l+3}} Y_{l+1,m+1/2}(\theta,\varphi) \end{pmatrix} \end{pmatrix} \quad (11)$$

and

$$\phi^{(n,j,m,(-1)^l)} = \frac{N_{n,j,m,(-1)^l}}{R^{3/2}} \begin{pmatrix} i\left[j_{l+1}(k_\pm r) + \frac{j_{l+1}(k_\pm R)}{i_{l+1}(\lambda_\pm(k)R)} i_{l+1}(\lambda_\pm(k)r)\right] \begin{pmatrix} \sqrt{\frac{l-m+3/2}{2l+1}} Y_{l+1,m-1/2}(\theta,\varphi) \\ -\sqrt{\frac{l+m+3/2}{2l+3}} Y_{l+1,m+1/2}(\theta,\varphi) \end{pmatrix} \\ -\rho_\pm(k_\pm)\left[j_l(k_\pm r) + \frac{j_l(k_\pm R)}{i_l(\lambda_\pm(k)R)} i_l(\lambda_\pm(k)r)\right] \begin{pmatrix} \sqrt{\frac{l+m+1/2}{2l+1}} Y_{l,m-1/2}(\theta,\varphi) \\ \sqrt{\frac{l-m+1/2}{2l+1}} Y_{l,m+1/2}(\theta,\varphi) \end{pmatrix} \end{pmatrix}, \quad (12)$$

with eigenenergies of the electrons (+) and holes (-) given by:

$$E_\pm(k) = \frac{1}{2}\left[\gamma k^2 \pm \sqrt{\left(E_g^{bulk} + \alpha k^2\right)^2 + \beta^2 k^2}\right], \quad (13)$$

where

$$\lambda_\pm(k) = \sqrt{\frac{2\alpha E_g^{bulk} + \beta^2 + (\alpha^2 - \gamma^2)k^2 + 4\gamma E_\pm(k)}{\alpha^2 - \gamma^2}}, \quad (14)$$

In the above equations, $\alpha = \hbar^2/2m_h + \hbar^2/2m_e$, $\beta = 2\hbar P/m_0$, $\gamma = \hbar^2/2m_h - \hbar^2/2m_e$. $j_l(x)$ and $i_l(x)$ are the spherical Bessel and modified spherical Bessel functions of the first kind, respectively, and $Y_{l,m}(\theta,\varphi)$ are the spherical harmonics.[38]

The quantum numbers $n = [1, 2 \cdots]$, $j = l + \frac{1}{2}$ ($l = [0, 1, \cdots]$), $m = [-j \cdots j]$ and $\pi = \pm 1$ correspond to the energy level, total angular momentum, projection of the angular momentum and parity, respectively. The value of $k$ is given by the $n^{th}$ lowest positive solution of the equations

$$\rho_\pm(k) j_{l+1}(ka) i_l(\lambda_\pm a) - \mu_\pm(k) j_l(ka) i_{l+1}(\lambda_\pm a) = 0 \quad (15)$$

for $\pi = (-1)^l$ or

$$\rho_\pm(k) j_l(ka) i_{l+1}(\lambda_\pm a) + \mu_\pm(k) j_{l+1}(ka) i_l(\lambda_\pm a) = 0 \quad (16)$$

for $\pi = (-1)^{l+1}$, and $\rho_\pm(k)$ and $\mu_\pm(k)$ are defined as:

$$\rho_\pm(k) = \frac{1}{\beta k}\left[\frac{E_g}{2} + (\alpha + \gamma)k^2 - 2E_\pm(k)\right]$$
$$\mu_\pm(k) = \frac{1}{\beta \lambda_\pm(k)}\left[E_g - (\alpha + \gamma)\lambda_\pm(k)^2 - 2E_\pm(k)\right]. \quad (17)$$

The normalization is given by:

$$\frac{1}{N_{n,j,m,(-1)^{l+1}}^2} = \int d\Omega \int_{x=0}^1 x^2 dx \left(\left\{\rho_\pm(k_\pm)\left[j_{l+1}(k_\pm Rx) + \frac{j_{l+1}(k_\pm R)}{i_{l+1}(\lambda_\pm(k)R)} i_{l+1}(\lambda_\pm(k)Rx)\right]\right\}^2 + \left[j_l(k_\pm Rx) + \frac{j_l(k_\pm R)}{i_l(\lambda_\pm(k)R)} i_l(\lambda_\pm(k)Rx)\right]^2\right)$$

$$\frac{1}{N_{n,j,m,(-1)^l}^2} = \int d\Omega \int_{x=0}^1 x^2 dx \left(\left\{\rho_\pm(k_\pm)\left[j_l(k_\pm ax) + \frac{j_l(k_\pm a)}{i_l(\lambda_\pm(k)a)} i_l(\lambda_\pm(k)ax)\right]\right\}^2 + \left[j_{l+1}(k_\pm ax) + \frac{j_{l+1}(k_\pm a)}{i_{l+1}(\lambda_\pm(k)a)} i_{l+1}(\lambda_\pm(k)ax)\right]^2\right). \quad (18)$$

The above wave functions and energies are then used to evaluate numerically the matrix elements of $V_{rsut}$ given by Eq. (4), the rate $\Gamma_S$, and the number of excitons generated upon excitation (cf., Eq. (7)).

## IV. RESULTS

### A. The role of the effective mass

To quantify the role of the effective masses of the charge carriers on the efficiency of MEG, we have preformed a set of calculations for two electron effective masses: $m_e/m_0 = \frac{1}{6}, 2$ and for a range of hole effective masses: $m_h/m_0 = \frac{1}{6}, \frac{2}{5}, \frac{2}{3}, 2$. Other model parameters are based on the PbS parameters of Table 1. The phonon decay rate was taken as $\gamma = 1\text{ps}^{-1}$. The results for a QD of 10nm diameter are shown in Figure 1 where we plot the average number of excitons generated as a function of the excitation energy (in scaled units $E/E_g$ where $E_g$ is the fundamental gap of the NC) for different effective masses. Left and right panels display the results for a light and heavy electron, respectively. The results are averaged over a 5% size distribution of the nanocrystals and an energy window of $\approx \pm \frac{1}{4} E_g$.

The general trends and the conclusions that can be drawn are quite clear. We find that when the effective masses of the two carriers are quite similar, MEG efficiencies are larger compared to the case where the two masses differ significantly, at the energy range shown. Moreover, the onset of MEG is below $3E_g$ even when the two masses are equal as a result of band mixing. Highest MEG efficiencies occur when the effective masses of both carriers are small and similar. Since the results are based on Fermi's golden rule, which breaks down for very small effective mass due to the decrease in the density of states, there is a lower bound on the magnitude of the effective mass. The behavior seen in Figure 1 holds qualitatively for different phonon emission rates or for other NC sizes, varied within an experimentally relevant range.

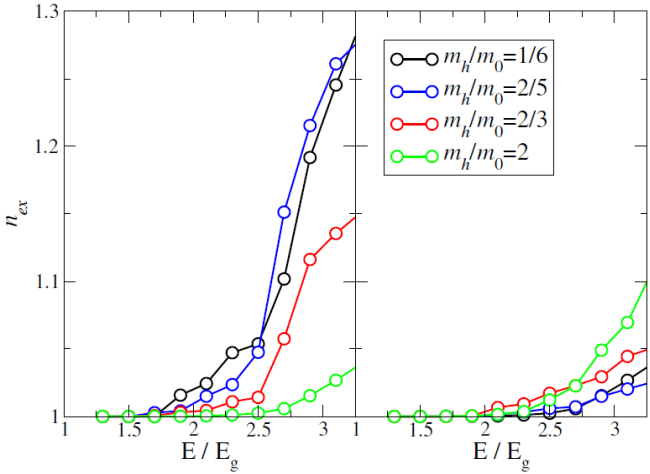

Figure 1: Number of excitons per photon, $n_{ex}$, as a function of scaled energy ($E/E_g$) for various values of the hole effective masses. Left panel shows the results for an electron effective mass of $m_e/m_0 = 1/6$ and the right panel shows results for an electron effective mass of $m_e/m_0 = 2$. The remaining parameters are for PbS nanocrystal with a diameter of $10 \pm 0.5$ nm. The results are averaged over an energy window of $\approx \pm \frac{1}{4} E_g$.

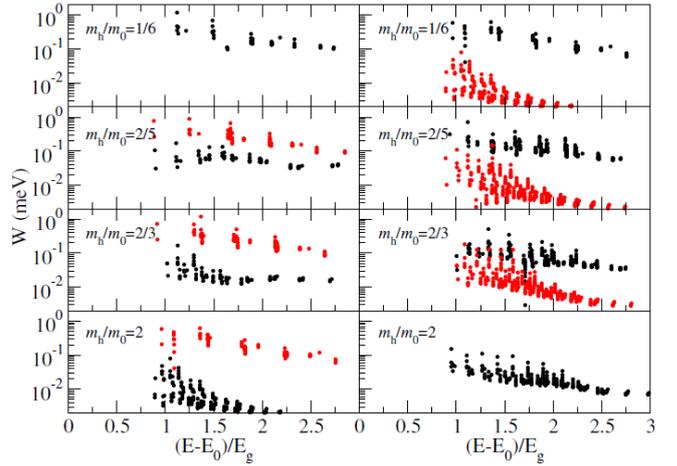

Figure 2: The average Coulomb coupling for electrons (red symbols) and holes (black symbols) for a 10nm diameter nanocrystal as a function of the scaled energy. The different panels represent different values of the hole effective mass. Left and right panels are for $m_e/m_0 = 1/6$ and $m_e/m_0 = 2$, respectively.

In Figure 2 we plot the average value of the Coulomb couplings, $\langle W_a \rangle = \sqrt{\frac{\hbar}{2\pi} \Gamma_a^- / \rho_T^-}$ for the electron decay, where $\Gamma_a^-$ is the negative trion formation rate and $\rho_T^-$ is the corresponding DOTS.[35] A similar expression, with $a \to i$ and $- \to +$, holds for the holes. Red and black symbols correspond to the average value $\langle W \rangle$ for electrons and holes in a given initial state, respectively. We first analyze the case when the effective mass of the electron equals that of the hole (upper left and lower right panels of Figure 2). Due to the symmetric band structure assumed by the model, the results for electrons and holes coincide in each case separately. At a given energy, $\langle W \rangle$ increases as the effective mass decreases since the corresponding wavefunctions are less oscillatory. This is evident comparing the results for $m_e/m_0 = m_h/m_0 = 2, \frac{1}{6}$. In the calculation of MEG efficiencies, the larger value of $\langle W \rangle$ for smaller masses overtakes the decrease in the DOTS (not shown here), leading to an overall increase of MEG efficiencies for small effective mass (see Figure 1).

When the two masses differ, $\langle W \rangle$ shows two distinct regimes with a *smaller* value for the *heavier* particle. In this case, the lighter particle can take most of the excitation energy and also, as explained above, assumes a larger value for $\langle W \rangle$ (see for example, the lower left panel in Figure 2). Comparing the trion formation rates of the lighter particle we find that they are only slightly influenced by an increase of the mass of the heavier particle. This, apparently, suggests that MEG would favor a large ratio of the effective masses, in contrast to the results shown in Figure 1.

This apparent paradox can be rationalized as follows. It turns out that the intuitive assumption that the lighter particle takes most of the excitation energy is, in fact, incorrect. Indeed, transitions where the lighter particle takes the excess energy are much stronger than other transitions. However, the density of singly excited states, where the heavy particle takes the excess energy, is much larger. Thus, the effective oscillator strength of such transitions is larger, often by two orders of magnitude. Since the average Coulomb coupling of the heavier particle is significantly lower, the overall efficiency decreases when the two masses differ, consistent with the picture shown in Figure 1.

### B. The role of band mixing

In order to test the effects of the coupling strength between the conduction and valence bands in the 4-band model, we have repeated the calculations by artificially changing the value of $P$ in the Hamiltonian given by Eq. (9). In Figure 3: Number of excitons per photon, $n_{ex}$, as a function of scaled energy for various values of the band coupling strength, $P$ (relative to the value of $P$ for PbS). The remaining parameters are taken for a 10nm diameter PbS nanocrystal. The results are averaged over an energy window of $\approx \pm \frac{1}{4} E_g$. we plot the MEG efficiencies for different values of $P$. MEG efficiencies vanish with diminishing band-couplings, consistent with the fact that for a 2-band model ($P \to 0$) the Coulomb coupling elements, $V_{rsut}$, are zero.

When $P$ is very large, the confinement energies increase relative to the bulk band gap, and thus, the density of states decreases, leading to very small MEG efficiencies. As intermediate values of $P$, MEG shows a maximal efficiency. For the present case, this occurs for values which are close to those of PbS.

### C. The role of phonon emission rate

The density of trion states increases rapidly with increasing excitation energies, leading to a rapid increase in the MEG

rates. Thus, one would expect that different phonon emission rates will only shift the onset of MEG. If, however, the increase of MEG rates near values comparable to the phonon emission rate is rather slow, then the changes in phonon emission rate may significantly affect the efficiency of MEG.

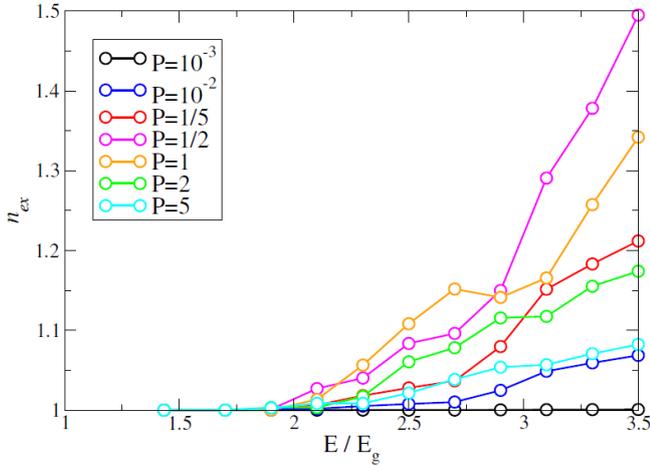

Figure 3: Number of excitons per photon, $n_{ex}$, as a function of scaled energy for various values of the band coupling strength, $P$ (relative to the value of $P$ for PbS). The remaining parameters are taken for a 10nm diameter PbS nanocrystal. The results are averaged over an energy window of $\approx \pm \frac{1}{4} E_g$.

In Figure 4: Number of excitons per photon, $n_{ex}$, as a function of the scaled energy for various phonon emission rates, calculated for a 10nm diameter PbS (left) and PbSe (right) nanocrystals. we plot the efficiency of MEG for various values of the phonon emission rates for PbS and PbSe NCs with a diameter of 10nm. The change in the phonon emission rate does not shift the onset of MEG, but rather changes the overall MEG efficiency. Even when the phonon emission rate increases by just a factor of 2, it leads to a decrease of the MEG efficiency by a similar factor. Comparing the results for PbS and PbSe, it is clear that differences observed in the MEG efficiencies results from the differences in the phonon emission rate and not from the differences in effective masses and band gaps, consistent with recent experimental reports.[18]

## V. CONCLUSIONS

In this paper we addressed the role of the effective mass of the electron and hole on MEG. We showed that when the two masses are *equal* and *small* the MEG efficiencies were maximized, consistent with high experimental MEG efficiencies for PbS. This is a result of the rapid increase in the Coulomb coupling relative to the slower decrease in the DOTS when the effective mass is reduced. Moreover, when the effective mass of the electron and hole are significantly different, as a result of asymmetric excitations allowed by band mixing, the MEG efficiencies are reduce, since excess energy given to the heavier particle does not contribute to the formation of multiexcitons.

We have also studied the role of couplings between the bands and the impact of the phonon emission rate on MEG. The former shows a maximum value for MEG efficiencies near coupling values of PbS. Variations in the phonon emission rate lead to an overall change in the MEG efficiencies, rather than shifting the onset of MEG. The differences observed experimentally between PbS and PbSe can be attributed to the difference in the phonon emission rate. For equal phonon emission rates, the two show similar MEG efficiencies, despite having different electron and hole model parameters.

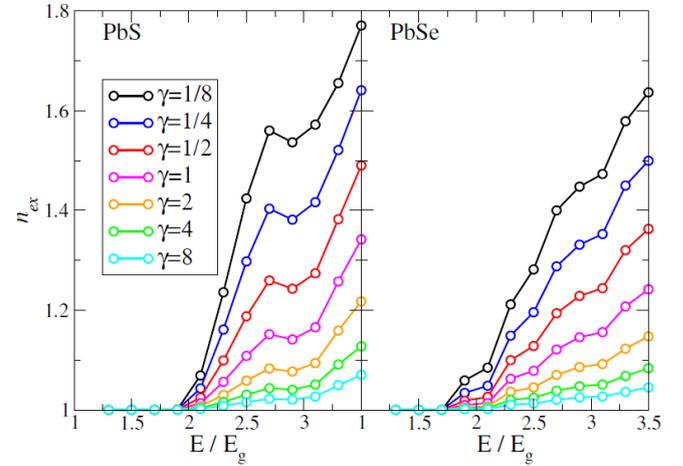

Figure 4: Number of excitons per photon, $n_{ex}$, as a function of the scaled energy for various phonon emission rates, calculated for a 10nm diameter PbS (left) and PbSe (right) nanocrystals. The results are averaged over an energy window of $\approx \pm \frac{1}{4} E_g$.

## Acknowledgments
This research was supported by the Israel Science Foundation (grant numbers 611/11, 1020/10) and by the FP7 Marie Curie IOF project HJSC. ER would like to thank the Center for Re-Defining Photovoltaic Efficiency Through Molecule Scale Control, an Energy Frontier Research Center funded by the U.S. Department of Energy, Office of Science, Office of Basic Energy Sciences under Award Number DE-SC0001085.